\documentclass[12pt]{article}

\newcommand{\nit}{\noindent}
\newcommand{\nl}{\newline}
\newcommand{\np}{\newpage}
\newcommand{\dsp}{\displaystyle}
\newcommand{\vs}[1]{\vspace{#1 ex}}
\newcommand{\hs}[1]{\hspace{#1 em}}
\newcommand{\bfr}{\begin{flushright}}
\newcommand{\efr}{\end{flushright}}
\newcommand{\bc}{\begin{center}}
\newcommand{\ec}{\end{center}}
\newcommand{\ben}{\begin{enumerate}}
\newcommand{\een}{\end{enumerate}}

\newcommand{\be}{\begin{equation}}
\newcommand{\ee}{\end{equation}}
\newcommand{\ba}{\begin{array}}
\newcommand{\ea}{\end{array}}
\newcommand{\ct}{\cite}
\newcommand{\bit}{\bibitem}
\newcommand{\dd}[2]{\frac{\partial{#1}}{\partial{#2}}}
\newcommand{\ag}{\alpha}
\newcommand{\bg}{\beta}
\newcommand{\gam}{\gamma}
\newcommand{\del}{\delta}
\newcommand{\eps}{\epsilon}
\newcommand{\ve}{\varepsilon}

\newcommand{\thg}{\theta}
\newcommand{\kg}{\kappa}
\newcommand{\lb}{\lambda}

\newcommand{\rg}{\rho}

\newcommand{\ps}{\psi}
\newcommand{\vf}{\varphi}
\newcommand{\og}{\omega}
\newcommand{\Gam}{\Gamma}

\newcommand{\Fg}{\Phi}

\newcommand{\Ps}{\Psi} 
\newcommand{\Og}{\Omega}
\newcommand{\Lb}{\Lambda}
\newcommand{\bgo}{\bar{\og}}
\newcommand{\bgl}{\bar{\lb}}
\newcommand{\bps}{\bar{\ps}}

\newcommand{\bFg}{\bar{\Fg}}
\newcommand{\bLb}{\bar{\Lb}}
\newcommand{\bPs}{\bar{\Ps}}
\newcommand{\bOg}{\bar{\Og}} 

\newcommand{\bv}{\bar{v}} 
\newcommand{\bz}{\bar{z}} 

\newcommand{\bE}{\bar{E}}
\newcommand{\bF}{\bar{F}}
\newcommand{\bH}{\bar{H}}

\newcommand{\bM}{\bar{M}}
\newcommand{\bN}{\bar{N}}
\newcommand{\bR}{\bar{R}}

\newcommand{\cD}{{\cal D}}
\newcommand{\cF}{{\cal F}}
\newcommand{\cH}{{\cal H}} 
\newcommand{\cL}{{\cal L}}

\newcommand{\lh}{\left(}
\newcommand{\rh}{\right)}
\newcommand{\ld}{\left.}

\newcommand{\slashed}{\hspace{-1.1ex}/}
\newcommand{\Slashed}{\hspace{-1.6ex}/\hspace{.3ex}}
\newcommand{\der}{\partial}
\newcommand{\Der}{D}
\newcommand{\sDer}{\Der\Slashed}
\newcommand{\sder}{\der\slashed}

\begin{document}

\pagestyle{empty}
\begin{flushright}
NIKHEF/2003-008\\
UVIC-TH/03-07
\end{flushright} 
\vs{2} 

\begin{center}
{\Large{\bf{Relativistic fluid mechanics,}}} \\ 
\vs{3} 
{\Large{\bf{K\"{a}hler manifolds and supersymmetry}}}\\
\vs{5}

{\large T.S.\ Nyawelo$^*$ and J.W.\ van Holten$^{**}$} \\
\vs{2}

NIKHEF, Amsterdam NL \\
PO Box 41882 \\ 
1009 DB Amsterdam \\
The Netherlands \\
\vs{3}

{\large S.\ Groot Nibbelink$^{\dagger}$} \\
\vs{1.5} 

(CITA National Fellow) \\
\vs{1.5} 
Univ.\ of Victoria, Dept.\ of Physics and Astronomy \\ 
PO Box 3055 STN CSC \\
Victoria BC, V8W 3P6 \\
Canada \\
\vs{4}

{\small{ \bf{Abstract} }} \\
\end{center}

\nit
{\small{We propose an alternative for the Clebsch decomposition of 
currents in fluid mechanics, in terms of complex potentials taking values in 
a K\"{a}hler manifold. We reformulate classical relativistic fluid mechanics
in terms of these complex potentials and rederive the existence of an 
infinite set of conserved currents. We perform a canonical analysis to 
find the explicit form of the algebra of conserved charges. 
The K\"{a}hler-space formulation of the theory has a natural supersymmetric 
extension in 4-$D$ space-time. It contains a conserved current, but also a
number of additional fields complicating the interpretation. Nevertheless,
we show that an infinite set of conserved currents emerges in the vacuum
sector of the additional fields. This sector can therefore be identified with 
a regime of supersymmetric fluid mechanics. Explicit expressions for 
the current and the density are obtained. }}

\vfill
\footnoterule 
\nit
{\footnotesize{
\begin{tabular}{rp{14cm}}
$^{*}$& {\tt e-mail:\  tinosn@nikhef.nl} \\
$^{**}$& {\tt e-mail:\  v.holten@nikhef.nl} \\
${}^{\dagger}$ &  {\tt e-mail:\  grootnib@uvic.ca} \\
& 
(Address after August 22nd: 
School of Physics \& Astronomy, 
University of Minnesota, 116 Church Street S.E.,
Minneapolis, MN 55455, USA.)
\end{tabular}
}}


\np
\pagestyle{plain} 
\pagenumbering{arabic}

\section{Introduction} 

Relativistic fluid mechanics has applications in the laboratory, e.g.\ in plasma 
physics and heavy ion collisions, as well as in astrophysics and cosmology 
\ct{ll,weinberg}. In recent times various extensions and reformulations 
of the theory have been proposed. The inclusion of non-zero vorticity and 
the role of non-trivial vortex topology in passing from the hamiltonian to the
lagrangean description have been studied, e.g.\ in refs.\ \ct{carter}-\ct{jackiw}. 
Non-abelian extensions of the theory have been found in 
\ct{jack-nair-pi}-\ct{jackiw2}, and supersymmetric models of fluid dynamics have 
been proposed in \ct{hoppe}-\ct{das-pop}. The role of space and space-time 
symmetries has been investigated in \ct{oraif-sreed,hass-horv} and references 
therein. A rather remarkable result is the existence of an infinite set of conserved 
currents in 4-$D$ space-time, related to the reparametrization invariance in the 
space of potentials \ct{bhsv,gosh}. This seems to offer an important key to identifying
fluid-dynamical phases of 4-$D$ relativistic field theory. In this paper we use the 
existence of such an infinite system of currents to find a fluid-dynamical
regime in a class of supersymmetric models. 

This paper is organized as follows. In sect.\ 2 we first recall the basic facts 
about non-dissipative  relativistic fluid mechanics. We propose an alternative 
to the standard Clebsch parametrization, based on complex potentials
taking values in a K\"{a}hler manifold, and we rederive the fluid equations 
in this formalism. In sect.\ 3 we show the existence of a topological invariant 
(the vortex linking number), and an infinite set of divergence-free currents. 
This is followed by a discussion of the canonical structure of the theory in terms 
of Dirac-Poisson brackets in sect.\ 4. We also compute the algebra of the 
conserved charges. The constructions are illustrated with a simple model for 
a fluid with an extremal value of the state parameter $\eta = p/\ve = 1$.
In sect.\ 5 we introduce 4-$D$ supersymmetry into the structure, by showing
how the potentials and the fluid current can be naturally incorporated in
$N = 1$ superfields. We propose a superfield action and present its component 
form, which is a generalization of the model proposed in ref.\ \ct{hngn}. In sect.\ 6 
we study currents and their conservation laws in the supersymmetric model. 
We show that there exists a regime in which an infinite number of currents is
reobtained; this regime we interpret as the description of a supersymmetric fluid. 
We finish with a discussion of our results and possible extensions.

\section{Relativistic fluid mechanics \label{s1}}

The equations of motion of a perfect (dissipationless) relativistic 
fluid can be expressed in terms of a conserved and symmetric 
energy-momentum tensor $T_{\mu\nu}$, derived from Poincar\'{e} 
invariance by Noether's theorem. 
The general form of the energy-momentum tensor of a relativistic 
perfect fluid is (see, e.g.\ \ct{ll,weinberg}):
\be
T_{\mu\nu} = p g_{\mu\nu} + (\ve + p) u_{\mu} u_{\nu},
\label{1.1}
\ee 
where $p$ is the pressure, $\ve$ is the energy-density and 
$u^{\mu}$ is the velocity four-vector, which in natural units ($c = 1)$
is a time-like unit vector: $u_{\mu}^2 = -1$. Local 
energy-momentum conservation is expressed by the vanishing of the 
four-divergence of the energy-momentum tensor
\be 
\der^{\mu} T_{\mu\nu} = 0. 
\label{1.2}
\ee 
The conserved energy-momentum four-vector is then given in a
laboratory inertial frame by 
\be 
P_{\mu} = \int_{t = t_0} d^3x\, T_{\mu 0}, \hs{2} 
 \frac{dP_{\mu}}{dt} = 0.
\label{1.3}
\ee 
In addition to the conservation of energy and momentum, the 
fluid density is conserved during ordinary flow as well. 
This is expressed by the vanishing divergence of the fluid 
density current $j^{\mu}$:
\be 
\der_{\mu} j^{\mu} = 0, \hs{2} j^{\mu} = \rg u^{\mu},
\label{1.4}
\ee 
where $\rg$ represents the local fluid density in the local instantaneous rest 
frame; the normalization of the four velocity then implies that the current 
satisfies 
\be 
- j_{\mu}^2 = \rg^2 \geq 0.
\label{1.5}
\ee  
Thus the local fluid density is defined in a Lorentz-invariant manner.
In a space-plus-time formulation, equation (\ref{1.4}) is seen
to imply the equation of continuity
\be 
\der \cdot j = \der_t(\rg \gam) + \nabla_i (\rg \gam v^i) = 0, 
\hs{2} \gam = \lh 1 - {\bf v}^2 \rh^{-1/2}.
\label{1.6}
\ee 
Because of the vanishing divergence, for general fluid flow the current 
has three independent components. A standard way to express this is to 
write the current in terms of three scalar potentials $(\thg, \ag, \bg)$;
they are introduced as lagrange multipliers combined in an auxiliary 
vector field $a_{\mu}$, with the Clebsch decomposition
\be 
a_{\mu} = \der_{\mu} \thg + \ag \der_{\mu} \bg.
\label{1.7}
\ee 
In this formalism the component $\thg$ describes pure potential flow, 
whilst $\ag$ and $\bg$ are necessary to include non-zero vorticity; 
for a review, see \ct{jackiw}.

In the present paper we advocate an alternative to the Clebsch 
decomposition, which is mathematically equivalent but has several 
advantages: it gives insight into the construction of an infinite 
set of conserved currents \ct{bhsv,gosh}, and it allows a straightforward 
supersymmetric generalization. 

Our approach consists in replacing the real Clebsch potentials 
$(\thg,\ag,\bg)$ by one real potential $\thg$ and one complex potential 
$z$, with its conjugate $\bz$. In terms of these potentials we propose 
a general Lagrange density for a relativistic fluid, reproducing the 
conserved energy-momentum tensor (\ref{1.1}). It is  given by the expression 
\be 
\ba{lll}
\cL[j^{\mu},\thg,\bz,z] & = & - j^{\mu} a_{\mu} - f(\rg) \\
 & & \\
 & = & -j^{\mu} \lh \der_{\mu} \thg 
 + i K_z \der_{\mu} z - i K_{\bz} \der_{\mu} \bz\rh - f\lh \sqrt{-j^2} \rh. 
\ea
\label{1.8}
\ee 
Here $K(z,\bz)$ is a real function of the complex potentials, 
which we refer to as the K\"{a}hler potential, $K_z$ and $K_{\bz}$
are its partial derivatives w.r.t.\ $z$ and $\bz$, and $f$ is a
function of $\rg = \sqrt{-j^2}$ only. 

The equations motion derived from (\ref{1.8}) are 
\be 
\ba{l}
\dsp{ f^{\prime}\frac{j_{\mu}}{\sqrt{-j^2}} = \der_{\mu} \thg +  
 i K_z \der_{\mu} z - i K_{\bz} \der_{\mu} \bz, \hs{2} \der \cdot j = 0, }\\
 \\ 
-2 i K_{z\bz}\, j \cdot \der z = 2 iK_{z\bz}\, j \cdot \der \bz = 0.
\ea 
\label{1.9} 
\ee 
Translation invariance of the action constructed from $\cL$ implies 
the conservation of the energy-momentum tensor 
\be 
T_{\mu\nu} = g_{\mu\nu} \lh f^{\prime} \sqrt{-j^2} - f \rh 
 + f^{\prime}\, \frac{j_{\mu}j_{\nu}}{\sqrt{-j^2}}, \hs{2} 
 \der^{\mu} T_{\mu\nu} = 0, 
\label{1.10}
\ee 
where $f^{\prime}$ is the derivative of $f(\rg)$ w.r.t.\
its argument $\rg = \sqrt{-j^2}$. Writing $j_{\mu} = \rg u_{\mu}$, 
this energy-momentum tensor is of the form (\ref{1.1}) with 
\be 
\ve = f(\rg), \hs{2} 
p = \rg f^{\prime}(\rg) - f(\rg).
\label{1.11}
\ee 
Hence the pressure is the negative of the Legendre transform of the specific 
energy w.r.t.\ the density $\rg$. Observe, that linear relations between 
pressure and specific energy correspond to power-law specific energies:
\be 
\ve = f(\rg) = \ag \rg^{(1 + \eta)} \hs{1} \Rightarrow \hs{1} 
 p = \eta \ve. 
\label{1.12}
\ee 

\section{Conservation laws} 

The essential elements of the class of fluid-dynamical models presented 
above are the existence of a divergence-free density current $j_{\mu}$ 
and a divergence-free energy-momentum tensor $T_{\mu\nu}$. We now 
show that there exist still other divergence-free currents, connected with 
conserved charges in the models defined above. 

First we recall the construction of a conserved topological charge, 
related to the linking number of vortices. Following Carter \ct{carter} 
we define the momentum density 
\be 
\pi_{\mu} = \ld \frac{\del \cL}{\del u^{\mu}}\right|_{\rg} = 
 \rg \lh \der_{\mu} \thg + i K_z \der_{\mu} z -  
 i K_{\bz} \der_{\mu} \bz \rh.
\label{2.1}
\ee 
Observe that the auxiliary vector potential $a_{\mu}$ is related to the 
momentum density by $\pi_{\mu} = \rg a_{\mu}$: 
\be 
\ba{lll} 
a_{\mu} & = & \dsp{ \der_{\mu} \thg + i K_z \der_{\mu} z - i K_{\bz} 
 \der_{\mu} \bz = f^{\prime}\, \frac{j_{\mu}}{\sqrt{-j^2}} 
 = f^{\prime} u_{\mu}, }\\
 & & \\ 
\pi_{\mu} & = & \dsp{  \rg f^{\prime} u_{\mu} 
 = (p + \ve) u_{\mu}. }
\ea 
\label{2.2}
\ee
From the definition it follows, that the axial current defined by $k^{\mu} = 
\ve^{\mu\nu\kg\lb}\, a_{\nu} \der_{\kg} a_{\lb}$ is divergence-free:
\be 
\der_{\mu} k^{\mu} = \ve^{\mu\nu\kg\lb}\, \der_{\mu} a_{\nu}\, 
 \der_{\kg} a_{\lb} = 0. 
\label{2.3}
\ee 
The conserved charge 
\be 
\og = \int d^3x\, k^0 = \int d^3x\, \ve^{ijk}\, a_i \der_j a_k,
\label{2.4}
\ee 
is a topological quantity (the linking number of vortices), given
by a pure surface term 
\be 
\og = \int d^3x\, \der_i \left[ i \ve^{ijk} \thg\, \der_j (K_{\bz}\der_k \bz 
 - K_{z} \der_k z) \right]
 =  - 2i \int d^3x\, \der_i \left[\ve^{ijk} \thg\, K_{\bz z} 
 \der_j \bz\, \der_k z \right]. 
\label{2.5}
\ee 
Next we show that there is an infinite set of conserved charges related 
to the reparametrization of the potentials \ct{bhsv}. As a first step 
observe that whenever $K_{z\bz} \neq 0$, the equations of motion for
the complex potentials $z$ and $\bz$ reduce to
\be
j \cdot \der z = 0, \hs{2} j \cdot \der \bz = 0. 
\label{2.6}
\ee 
It follows, that any current 
\be
J_{\mu}[G] = - 2 G(\bz,z) j_{\mu},
\label{2.7}
\ee 
is divergence-free:
\be 
\der \cdot J[G] = - 2\Bigl(G_z j \cdot \der z + G_{\bz} j \cdot \der \bz\Bigl) = 0.
\label{2.8}
\ee
which allows the construction of infinitely many conserved charges of the form 
\be 
Q[G] = \int d^3x\, J^0[G]. 
\label{2.9}
\ee 
The non-singularity of the K\"{a}hler potential is satisfied in all cases 
where $K_{z\bz}$ is the metric of a geodesically complete complex manifold. The 
simplest example is the complex plane with $K(z, \bz) = \bz z$ and hence $K_{z\bz} 
= 1$. Another example is the sphere $S^2 = CP^1$, with the K\"{a}hler 
potentials $K^{(\pm)}(z_{\pm},\bz_{\pm}) = \ln (1 + \bz_{\pm} z_{\pm} )$ to be used 
on the northern and southern hemisphere, respectively, related up to the real part 
of a holomorphic function by the analytic co-ordinate transformation $z_- = 1/z_+$; 
in this case 
\be 
K^{(\pm)}_{z\bz} = \frac{1}{(1 + \bz_{\pm} z_{\pm})^2}. 
\label{2.10}
\ee 

\section{Canonical structure} 

We now pass to the canonical formulation of the theory. First we 
define the canonical momenta 
\be 
\pi_{\thg} = \dd{\cL}{\dot{\thg}} = j_0, \hs{2} 
\pi_z = \dd{\cL}{\dot{z}} =  i K_z j_0, \hs{2}
\pi_{\bz} = \dd{\cL}{\dot{\bz}} = -i K_{\bz} j_0.
\label{2.11}
\ee 
With $\rg = \sqrt{ \pi^2_{\thg} - {\bf j}^2 }$ the hamiltonian density and
spatial current components are   
\be 
\cH = \frac{f^{\prime}(\rg)}{\rg}\, {\bf j}^2 + f(\rg),
\hs{2} 
\frac{f^{\prime}(\rg)}{\rg}\, {\bf j} = \mbox{\boldmath{$\nabla$}} \thg +  
 iK_z \mbox{\boldmath{$\nabla$}} z - i K_{\bz} \mbox{\boldmath{$\nabla$}} \bz.
\label{2.12}
\ee 
Obviously, the  last two equations (\ref{2.11}) are second-class 
constraints, expressing $(\pi_z, \pi_{\bz})$ in terms of the other 
phase-space variables $(z, \bz, \pi_{\thg})$:
\be 
\chi_z = \pi_z - i K_z \pi_{\thg} = 0, \hs{2} 
\chi_{\bz} = \pi_{\bz} + i K_{\bz} \pi_{\thg} = 0. 
\label{2.13}
\ee 
To describe the canonical dynamics on the reduced phase-space determined 
by these equations, we introduce Poisson-Dirac brackets 
\be 
\left\{ A, B \right\}^* = \left\{ A, B \right\} - 
 \left\{ A, \chi_i \right\} C^{-1}_{ij} \left\{ \chi_j, B \right\},
\label{2.14}
\ee 
where $C^{-1}$ is the inverse of the matrix of constraint brackets
\be 
C_{ij} = \left\{ \chi_i({\bf r},t) , \chi_j({\bf r}^{\prime}, t) \right\} 
 = \lh \ba{cc} 0 & -2i K_{z\bz} \pi_{\thg} \\ 
               2i K_{z\bz} \pi_{\thg} & 0  \ea \rh\, 
 \del({\bf r} - {\bf r}^{\prime}).
\label{2.15}
\ee 
From the definition (\ref{2.14}) it follows, that in the reduced 
phase space spanned by $(z, \bz, \thg, \pi_{\thg})$ the canonical 
Poisson-Dirac brackets are 
\be 
\ba{ll}
\dsp{ \left\{ z({\bf r}, t), \bz({\bf r}^{\prime}, t) \right\}^* 
 = \frac{-i}{2K_{z\bz} \pi_{\thg}}\, \del({\bf r} - {\bf r}^{\prime}), }&
 \left\{ \thg({\bf r}, t), \pi_{\thg}({\bf r}^{\prime}, t) \right\}^* 
 = \del({\bf r} - {\bf r}^{\prime}), \\
 & \\
\dsp{ \left\{ z({\bf r}, t), \thg({\bf r}^{\prime}, t) \right\}^* 
 = \frac{K_{\bz}}{2K_{z\bz} \pi_{\thg}}\, \del({\bf r} - {\bf r}^{\prime}), 
 }&
\dsp{ \left\{ \bz({\bf r}^{\prime}, t),  \thg({\bf r}, t) \right\}^* 
 = \frac{K_{z}}{2K_{z\bz} \pi_{\thg}}\, \del({\bf r} - {\bf r}^{\prime}). 
 }
\ea
\label{2.16}
\ee 
With the help of these rules we can determine the algebra of the conserved 
charges. It is useful to revert to a geometrical notation in terms of a 
simple K\"{a}hler manifold with metric $g_{z\bz} = g_{\bz z} = K_{z\bz}$
and its inverse $g^{z\bz} = g^{\bz z} = 1/K_{z\bz}$. The action of the 
$Q[G]$ on the potentials then is:
\be
\ba{l} 
\dsp{ \del_G \thg = \left\{ Q[G], \thg \right\}^* 
 = 2 G - g^{z\bz} \lh K_{\bz} G_z + K_z G_{\bz} \rh, }\\
 \\
\dsp{ \del_G z = \left\{ Q[G], z \right\}^* = - i\, g^{z\bz} G_{\bz}, }\\
 \\
\dsp{ \del_G \bz = \left\{ Q[G], \bz \right\}^* = i\, g^{\bz z} G_z. }
\ea
\label{2.17}
\ee 
If $G(\bz,z)$ is taken to transform as a scalar on the complex manifold, the
transformations $\del_G$ are seen to take a covariant form and represent 
a reparametrization of the complex target manifold of the potentials $(\bz,z)$. 
These transformations have the property that they leave the auxiliary vector 
potential (one-form) invariant: 
\be 
a = dx^{\mu} a_{\mu} = d\thg + i K_z dz - i K_{\bz} d\bz \hs{1} 
 \Rightarrow \hs{1} \del_G a = 0.
\label{2.18}
\ee 
As $a_{\mu}^2 = - f^{\prime\, 2}(\rg)$, it follows that also $\del_G \rg = 0$\
and $\del_G j_{\mu} = 0$. It is clear, that the transformations 
$\del_G (\thg, z, \bz)$ in eq.(\ref{2.17}) together with $\del_G j_{\mu} = 0$ 
define an infinite set of global symmetries of the lagrangean (\ref{1.8}) and 
the hamiltonian (\ref{2.12}).: 
\be
\left\{ Q[G], H \right\}^* = 0, \hs{2} H = \int d^3x\, \cH. 
\label{2.18.1}
\ee 
These  symmetries imply the reparame\-trization invariance of the
equations of motion of the auxiliary vector potential. 
 
A particularly simple instance is the flat complex plane with $K = \bz z$. 
Then 
\be
\ba{l} 
\dsp{ \left\{ Q[G], \thg \right\}^* = 2\,G - \lh zG_z + \bz G_{\bz} \rh, }\\
 \\
\dsp{ \left\{ Q[G], z \right\}^* = - i\, G_{\bz}, \hs{2} 
            \left\{ Q[G], \bz \right\}^* = i\,  G_z. }
\ea
\label{2.19}
\ee 
For this special choice we then find 
\be 
\ba{l}
\left\{ Q[1], \thg \right\}^* = 2, \hs{2} \left\{ Q[K], \thg \right\}^* =  
0, \\
 \\
\dsp{ \left\{ Q[1], z \right\}^* = 0, \hs{2} \left\{ Q[K], z \right\}^* 
 = i\, z, }\\ 
 \\
\dsp{ \left\{ Q[1], \bz \right\}^* = 0, \hs{2} \left\{ Q[K], \bz \right\}^* 
 = - i\, \bz.} 
\ea
\label{2.20}
\ee 
Returning to the case of general $K(z,\bz)$, we finally notice the closure of 
the algebra of conserved charges:
\be 
\left\{ Q[G^{(1)}], Q[G^{(2)}] \right\}^* =  Q[G^{(3)}], 
\label{2.21}
\ee 
with 
\be 
G^{(3)} = i g^{z\bz} \lh G^{(1)}_{z} G^{(2)}_{\bz} - 
 G^{(1)}_{\bz} G^{(2)}_{z} \rh. 
\label{2.22}
\ee  
This expression has itself the structure of a Poisson bracket on the 2-d manifold spanned 
by $(\bz,z)$. 

An example of a simple model relevant for later discussions is defined by
\be 
f(\rg) = \frac{\lb}{2}\, \rg^2 \hs{1} \rightarrow \hs{1} 
p = \ve = \frac{\lb}{2}\, \rg^2. 
\label{2.23}
\ee 
The hamiltonian density then becomes 
\be 
\cH = \frac{\lb}{2} \lh {\bf j}^2 + \pi_{\thg}^2 \rh,
\label{2.24}
\ee
where the current is 
\be 
{\bf j} = \frac{1}{\lb}\, \lh \mbox{\boldmath{$\nabla$}} \thg + i K_z 
 \mbox{\boldmath{$\nabla$}} z - i K_{\bz} \mbox{\boldmath{$\nabla$}} \bz \rh. 
\label{2.25}
\ee 
In this case the brackets with the current components become
\be 
\ba{l} 
\left\{ z({\bf r},t), {\bf j}({\bf r}^{\prime},t) \right\}^* = \dsp{
 \frac{\mbox{\boldmath{$\nabla$}} z}{\lb\pi_{\thg}}\, \del({\bf r} - 
 {\bf r}^{\prime}), \hs{2} 
\left\{ \bz({\bf r},t), {\bf j}({\bf r}^{\prime},t) \right\}^* = 
 \frac{\mbox{\boldmath{$\nabla$}} \bz}{\lb\pi_{\thg}}\, \del({\bf r} - 
 {\bf r}^{\prime}), }\\ 
 \\ 
\left\{ \thg({\bf r},t), {\bf j}({\bf r}^{\prime},t) \right\}^* = \dsp{
 \frac{i}{\lb\pi_{\thg}}\, \lh K_{\bz} \mbox{\boldmath{$\nabla$}} \bz - 
 K_{z} \mbox{\boldmath{$\nabla$}} z \rh \del({\bf r} - {\bf r}^{\prime}).  }
\ea
\label{2.26}
\ee 
Upon the identification $j_0 = \pi_{\thg}$, the brackets of the fields $\Fg = (\thg, z, \bz)$ 
with the hamiltonian can easily be checked to reproduce the field equations (\ref{1.9}):
\be 
\dot{\Fg} = \left\{ \Fg, H \right\}^*.
\label{2.27}
\ee 

\section{Supersymmetry}

The decomposition of the auxiliary vector in terms of real and complex scalar potentials
has a natural supersymmetric extension in 4-d Minkowski space-time. This leads to a 
proposal for a supersymmetric version of relativistic fluid dynamics in 4-d space-time. 
The supersymmetric extension is obtained by identifying the current $j_{\mu}$ and the 
auxiliary vector $a_{\mu}$ with the vector components $V_{\mu}$ and $A_{\mu}$ of two 
real superfields $V$ and $A$, with a general superfield action of the form
\be 
S[V,A] = \int d^4x \cL[V,A] = 
 \int d^4x \int d^2 \thg_+ \int d^2 \thg_- \lh \frac{1}{4}\, VA - F(V) \rh.
\label{4.1}
\ee 
Here $\thg_{\pm}$ are the positive/negative chirality components of the spinor 
co-ordinates of superspace. In terms of the multiplets of components $V = 
(C,\psi,M,V_{\mu},\lb,D)$ and $A = (B,\chi,N,A_{\mu},\og,G)$ the component 
lagrangean reads
\begin{eqnarray}
\cL[V,A] & =& \dsp  CG\, +\, BD\, +\, 2\, (M \bN + N \bM)\, -\, V \cdot A\, -\, 
 \der C \cdot \der B -\, F^{\prime}(C)\, D  
\nonumber \\[2ex]
 &  -& {\dsp  \bgl_+ \chi_+ - \bgl_- \chi_- - \bgo_+ \ps_+ - \bgo_- \ps_- 
 -\, \frac{1}{2}\, \bps_+ \stackrel{\leftrightarrow}{\sder} \chi_- - 
 \frac{1}{2}\, \bar{\chi}_+ \stackrel{\leftrightarrow}{\sder} \ps_- }
 \label{4.2}
\\[2ex]
 & -&  \dsp{   F^{\prime\prime}(C) \left[ 2\, |M|^2 - \frac{1}{2}\, 
 V_{\mu}^2 - \frac{1}{2}\, (\der_{\mu} C)^2 - \bgl_+ \ps_+ - \bgl_- \ps_-  
 -\, \frac{1}{2}\, \bps_+ \stackrel{\leftrightarrow}{\sder} \ps_- \right]
 }
\nonumber \\[2ex] 
 & +& \dsp \frac{1}{2}\, F^{\prime\prime\prime}(C) \left[ M \bps_- 
 \ps_- + \bM \bps_+ \ps_+ - i \bps_+ V\Slashed \ps_- \right] -  
 \frac{1}{8}\, F^{\prime\prime\prime\prime}(C)\, \bps_+ \ps_+ 
 \bps_- \ps_-   
\nonumber 
\end{eqnarray}
The decomposition (\ref{1.8}), (\ref{1.9}) of a vector field in terms of potentials 
can be extended to supersymmetric models by writing the auxiliary vector 
superfield in terms of  two sets of complex chiral superfields 
$(\Lb, \bLb, \Fg, \bFg)$, as follows:
\be 
A = \Lb + \bLb + K(\bFg,\Fg),
\label{4.3}
\ee 
where $K(\bFg,\Fg)$ is a real function of its superfield arguments; below it will
become clear that its lowest bosonic component $K(\bz,z)$ is the K\"{a}hler potential 
for the complex potentials $(\bz,z)$. We wish to point out, that a slight refinement of 
the theory can be obtained by taking $\Lb$ to be defined in terms of a real 
(vector) superfield $W$ by
\be 
\Lb = \frac{1}{4}\, \bar{D}_+ D_+ W, \hs{2} \bLb = \frac{1}{4}\, \bar{D}_- D_- W,
\label{4.4}
\ee 
with $D_{\pm}$ the usual covariant superspace derivatives. 
However, the equations for the physical degrees of freedom are the same as in the
case of $\Lb$ being a fundamental superfield, except for one mass parameter which 
is forced to vanish in the case of the decomposition (\ref{4.3}). In this paper we
treat only this simpler case. 

We label the components of the chiral superfield potentials by $\Lb = (v, \vf, E)$ 
and $\Fg = (z, \eta, H)$. Then the components of the auxiliary superfield $A$ 
are replaced by the expressions 
\be
\ba{l}
B = v + \bv + K(\bz, z), \\ 
 \\
\chi_{+} =  \vf_{+} + K_{\bz} \eta_+, \hs{3}  
 \chi_- = \vf_- + K_{z} \eta_- \\
 \\ 
N = \dsp{ \frac{1}{2}\, \left[ E + K_{\bz} \lh \bH - \frac{1}{2}\, 
 \bar{\Gamma}_{\bz\bz}^{\;\;\;\;\bz} \bar{\eta}_+ \eta_+ \rh \right], \hs{1} 
\bN = \frac{1}{2}\, \left[ \bE + K_{z} \lh H - \frac{1}{2}\, 
 \Gamma_{zz}^{\;\;\;\;z} \bar{\eta}_- \eta_- \rh \right], }\\ 
 \\
A_{\mu} = \dsp{ - i \der_{\mu} \lh v - \bv \rh + i \lh K_{,z} \der_{\mu} z -
 K_{,\bz} \der_{\mu} \bz \rh - \frac{i}{2}\, g_{\bz z}\, \bar{\eta}_+ \gam_{\mu} \eta_-, } \\
 \\
\og_{+} = \dsp{ \frac{1}{2} g_{\bz z} \lh H - \frac{1}{2}\, 
 \Gam_{zz}^{\;\;\;\;z} \bar{\eta}_- \eta_- \rh \eta_+ - g_{\bz z} 
 \sder \bz\, \eta_-, }\\
 \\
\og_- = \dsp{ \frac{1}{2}\, g_{\bz z} \lh \bH - \frac{1}{2}\, 
 \bar{\Gam}_{\bz \bz}^{\;\;\;\;\bz} \bar{\eta}_+ \eta_+ \rh \eta_- - g_{\bz z} \sder z\, 
 \eta_+, }\\ 
  \\
G = \dsp{ -\, 2 g_{\bz z}\, \der \bz \cdot \der z - \frac{1}{2}\, 
 g_{\bz z}\, \bar{\eta}_+ \stackrel{\leftrightarrow}{\sDer} \eta_- - 
 \frac{1}{8}\, R_{\bz z\bz z}\, \bar{\eta}_+ \eta_+ \bar{\eta}_- \eta_- }\\ 
 \\ 
 \dsp{ \hs{2.2} +\, \frac{1}{2}\, g_{\bz z} \lh \bH - \frac{1}{2}\, 
 \bar{\Gam}_{\bz\bz}^{\;\;\;\;\bz} \bar{\eta}_+ \eta_+ \rh \lh H - \frac{1}{2}\, 
 \Gam_{zz}^{\;\;\;\;z} \bar{\eta}_- \eta_- \rh. } 
\ea
\label{4.5}
\ee  
Here $g$, $\Gam$ and $R$ denote the metric, connection and curvature constructed 
from the K\"{a}hler potential $K$, respectively; moreover, the covariant derivative
of the chiral spinors $\eta_{\pm}$ are defined by
\be 
\sDer \eta_+ = \sder \eta_+ + \sder \bz\, \bar{\Gam}_{\bz\bz}^{\;\;\;\;\bz} \eta_+, 
\hs{2} \sDer \eta_- = \sder \eta_- + \sder z\, \Gam_{zz}^{\;\;\;\;z} \eta_-. 
\label{4.6}
\ee 
Upon substitution of these expression into eq.(\ref{4.2}) and elimination of the 
auxiliary fields, we then find the effective component lagrangean 
\be 
\ba{lll}
\cL_{eff} & = & - V^{\mu} \lh \der_{\mu} \thg + i K_z \der_{\mu} z - i K_{\bz} \der_{\mu} \bz  
 - \frac{i}{2}\ g_{\bz z} \bar{\eta}_+ \gam_{\mu} \eta_- \rh \\
 & & \\
 & & - \frac{1}{2} F^{\prime\prime}(C) \left[ - V^2 + (\der C)^2 + \bps_+ 
 \stackrel{\leftrightarrow}{\sder} \ps_- \right] - \frac{i}{2} F^{\prime\prime\prime}(C)
 \bps_+ V\Slashed \ps_- \\
 & & \\
 & & -\, C \lh 2 g_{\bz z}\, \der \bz \cdot \der z 
 + \frac{1}{2} g_{\bz z}\, \bar{\eta}_+ \stackrel{\leftrightarrow}{\sDer} \eta_- 
 + \frac{1}{8} R_{\bz z\bz z} \bar{\eta}_+ \eta_+ \bar{\eta}_- \eta_- \rh \\
 & & \\
 & & -\, \frac{1}{4C}\, g_{\bz z}\, \bar{\eta}_+ \gam^{\mu} \eta_-\, \bps_+ \gam_{\mu} \ps_- 
 + g_{\bz z}\, \lh \bps_+ \sder \bz \eta_- + \bps_- \sder z \eta_+ \rh \\ 
 & & \\
 & & -\, \frac{1}{8}\, F^{\prime\prime\prime\prime}(C)\, \bps_+ \ps_+ \bps_- \ps_-.
\ea 
\label{4.7}
\ee 
To make the connection with fluid mechanics, cf.\ eq.(\ref{1.8}), we have 
introduced the notation $\thg  = 2\,$Im$\,v$. It is obvious, that in the absence 
of fermions $(\ps_{\pm} = \eta_{\pm} = 0)$ and for $C = 0$ we reobtain the 
lagrangean (\ref{1.8}) with 
\be 
f(\rg) = \frac{1}{2}\, F^{\prime\prime}(0) \rg^2.
\label{4.8}
\ee
This is of the type (\ref{2.23}) with $\lb = F^{\prime\prime}(0)$.  The scalar $C$ and 
the spinor fields $\ps$ and $\eta$ describe additional dynamical fields. 
Being the co-efficient of the kinetic terms of the fields $(\bz, z)$ and $\eta_{\pm}$,
physics requires the scalar field $C$ to be non-negative. This can easily 
be achieved, for example by replacing the real superfield $V$ by another real 
superfield $W$ such that $V = e^W$. Thus we can take the condition 
$C \geq 0$ for granted.  

A simpler version of this action with only a single real scalar potential $\thg$ (hence
$\bz = z = \eta_{\pm} = 0$) was discussed in \ct{hngn}. In the absence of the 
complex potentials  its bosonic reduction in the hydrodynamical regime describes 
only potential flow; therefore in this model vorticity arises only by the presence of 
fermions.

\section{Supersymmetric fluid dynamics}

The supersymmetric extension of the action for fluid dynamics constructed above
generally goes at the expense of most of the infinitely many conservation laws 
related to reparametrizing the  potential, eqs.(\ref{2.7}), (\ref{2.9}). This can already 
be inferred from the bosonic part of the theory. Consider the bosonic terms in the
equations of motion for the current and the potentials: 
\be 
\der \cdot V = 0, \hs{2} 
2 \cD \cdot (C \der z) - 2i\, V \cdot \der z = 2 \cD \cdot (C \der \bz) + 
2 i\, V \cdot \der \bz = 0,
\label{4.9}
\ee 
Here $D_\mu$ denotes a covariant derivative containing the K\"{a}hler connection, e.g.\
\be 
D_{\mu} (\der_{\nu} z) = \der_{\mu} \der_{\nu} z 
 + \Gam_{zz}^{\;\;\;z} \der_{\mu} z \der_{\nu} z.
\label{4.10}
\ee 
Now construct the currents 
\be 
J_{\mu}[G] = -2 G(\bz,z) V_{\mu} - 2 i C (G_{,z} \der_{\mu} z - G_{,\bz} \der_{\mu} \bz),
\label{4.11}
\ee 
where $G(\bz, z)$ is a real function of the complex scalar fields. 
Using eqs.(\ref{4.9}) it can be seen to satisfy 
\be 
\der \cdot J[G] = - 2 i C \lh G_{;z;z} (\der z)^2 - G_{;\bz;\bz} (\der \bz)^2 \rh.
\label{4.12} 
\ee
It follows that the divergence of the current vanishes identically only 
for functions $G(\bz,z)$ such that the homogeneous second derivatives 
w.r.t.\ $\bz$ and $z$ vanish:
\be 
G_{;z;z} = G_{;\bz;\bz} = 0.
\label{4.13}
\ee 
This holds if and only if the gradients of $G(\bz, z)$ represent holomorphic 
Killing vectors $(R(z), \bar{R}(\bz))$, generating isometries of the K\"{a}hler 
manifold; a proof is presented in appendix A. As the number of independent 
isometries of a finite-dimensional manifold is finite, no infinite set of conserved
currents can be generated by Killing vectors. 

Still, as anticipated an  infinite set of conserved currents
$J_{\mu}[G]$ is obtained for all models (\ref{4.7}) under the
restriction $C=0$. Therefore we identify the manifold  
of states with $C = 0$ as the hydrodynamical regime of the
supersymmetric models constructed here. 
\vspace{2ex}

\nit
In the fully supersymmetric case it is easy to prove that holomorphic Killing vectors 
generate conserved currents. It can largely be done directly in superspace, as 
follows: let $R(\Fg)$ and $\bR(\bFg)$ be holomorphic functions of the chiral superfields 
$\Fg$ and $\bFg$, respectively, defining Killing vectors of the K\"{a}hler metric 
$g_{\Fg\bFg} = K_{,\Fg \bFg}$;  the essential element is, that under these transformations
the K\"{a}hler potential transforms non-homogeneously into the real part of 
a holomorphic function: 
\be 
\del \Fg = R(\Fg), \hs{2} \del \bFg = \bR(\bFg), \hs{2} 
\del K(\bFg, \Fg) = \cF_R(\Fg)  + \bar{\cF}_R(\bFg). 
\label{4.17}
\ee 
The precise form of the holomorphic function $\cF_R(\Fg)$ depends on the Killing vector 
$R(\Fg)$. Then defining the superfield transformations 
\be 
\del V = 0, \hs{2} \del \Lb = - \cF_R(\Fg),  \hs{2} \del  \bLb = - \bar{\cF}_R(\bFg), 
\label{4.18}
\ee 
the superspace action (\ref{4.1}) with $A$ given by (\ref{4.3}) is seen to be invariant. 
Noether's theorem then guarantees the existence of a conserved current for each
of the Killing vectors. In components they take the form
\begin{eqnarray}
J_{\mu}[G] & = & \dsp{ - 2 G V_{\mu} 
 - 2i C G_{,z} \lh \der_{\mu} z - \frac{1}{2C} \bps_+ \gam_{\mu} \eta_- \rh 
 + 2 i C G_{,\bz} \lh \der_{\mu} \bz - \frac{1}{2C} \bps_- \gam_{\mu}
\eta_+ \rh }
\nonumber \\[2ex]
 & & \dsp{ -\, i C G_{;\bz;z}\, \bar{\eta}_- \gam_{\mu} \eta_+, }
\label{4.19}
\end{eqnarray}
where $G(\bz,z)$ is the Killing potential for the isometries $\del z = R(z)$ and their  
complex conjugates (see appendix A). For a generic real function $G(\bz,z)$ which
is not a Killing potential, the current $J_{\mu}[G]$  is not conserved, unless one takes 
the limit $(C, \eta) \rightarrow 0$, such that the spinor field  $\eta$ vanishes as fast as 
$C$. Solutions of the model with this property we interpret as a supersymmetric fluid. 
 
To analyse this regime, we rescale the fermion fields as follows
\be
\ps_{\pm} = \frac{1}{\sqrt{F^{\prime\prime}(C)}}\, \Ps_{\pm}, \hs{2} 
 \eta_{\pm} = C \Og_{\pm}. 
\label{4.20}
\ee 
Then the lagrangean (\ref{4.7}) becomes
\begin{eqnarray}
\cL_{eff} & = & - V^{\mu} \lh \der_{\mu} \thg + i K_z \der_{\mu} z - i K_{\bz} \der_{\mu} \bz  
 - \frac{i}{2}\ C^2 g_{\bz z} \bar{\Og}_+ \gam_{\mu} \Og_- + 
 \frac{iF^{\prime\prime\prime}(C)}{2F^{\prime\prime}(C)}\,
\bar{\Psi}_+ \gam_{\mu} \Psi_- \rh 
\nonumber \\[2ex]
 & & + \frac{1}{2} F^{\prime\prime}(C) \lh V_{\mu}^2 - (\der_{\mu} C)^2 \rh
 - \frac{1}{2}\, \bar{\Psi}_+ \stackrel{\leftrightarrow}{\sder} \Psi_- 
  -\, \frac{F^{\prime\prime\prime\prime}(C)}{8 \left[ F^{\prime\prime}(C)\right]^2}\, 
 \bar{\Psi}_+ \Psi_+ \bar{\Psi}_- \Psi_- 
\nonumber \\[2ex]
 & & -\, 2 C g_{\bz z}\, \der \bz \cdot \der z 
 - \frac{1}{2} C^3 g_{\bz z}\, \bar{\Og}_+ \stackrel{\leftrightarrow}{\sDer} \Og_- 
 + \frac{1}{8}\, C^5  R_{\bz z\bz z}\, \bar{\Og}_+ \Og_+ \bar{\Og}_-
\Og_-  
\label{5.3} \\[2ex]
 & & + g_{\bz z}\, \frac{C}{\sqrt{F^{\prime\prime}(C)}}\, 
 \lh \bar{\Psi}_+ \sder \bz \Og_- + \bar{\Psi}_- \sder z \Og_+ \rh 
  -\, g_{\bz z} \frac{C}{4F^{\prime\prime}(C)}\, \bar{\Og}_+ \gam^{\mu} \Og_-\, 
 \bar{\Psi}_+ \gam_{\mu} \Psi_-.
\nonumber
\end{eqnarray}
We observe that in the limit $C = 0$ divergent terms can be avoided, provided 
$F^{\prime\prime}(0) \neq 0$. Then we can always normalize $F(C)$ such that 
$F^{\prime\prime}(0) = 1$; with this choice the quadratic vector term and the 
kinetic term of the real scalar $C$ have the canonical normalization. 

Next we observe, that there exist many choices of the function $F(C)$ 
such that also the coefficients of the bilinear and quartic terms in $\Psi$
are finite. Indeed, any function such that the second derivative has the 
expansion
\be 
F^{\prime\prime}(C) = 1 + \lb_1 C + \lb_2 C^2 + {\cal O}(C^3)
\label{5.4}
\ee
satisfies the conditions 
\be 
F^{\prime\prime}(0) = 1, \hs{1} F^{\prime\prime\prime}(0) = \lb_1, \hs{1} 
 F^{\prime\prime\prime\prime}(0) = 2 \lb_2,
\label{5.5}
\ee
and makes the lagrangean finite in the hydrodynamical regime. Having
established the existence of regular configurations with $C = 0$, the
expression for the current in this regime becomes 
\be 
V_{\mu}[C = 0] = \der_{\mu} \thg + i K_z \der_{\mu} z - i K_{\bz} \der_{\mu} \bz
 + \frac{i \lb_1}{2}\, \bar{\Psi}_+ \gam_{\mu} \Psi_-.
\label{5.6}
\ee
The bosonic part has the standard decomposition for a fluid density current; 
the last term is a fermionic extension required by supersymmetry. 

\nit
Next we consider the energy-momentum tensor and the equation for $C$; again 
in this regime. For $C = 0$, the symmetric energy-momentum tensor and the 
equation for $C$ derived from  (\ref{5.3}) reduce to 
\begin{eqnarray}
\lb_1 V_{\mu}^2 & = & \dsp{ 4 g_{z\bz} \der \bz \cdot \der z + i \lh 2 \lb_2 - \lb_1^2 \rh 
 \bPs_+ V\Slashed \Ps_- - 2 g_{z\bz} \lh \bPs_+ \sder \bz \Og_- +
\bPs_- \sder z \Og_+ \rh  }
\nonumber \\[2ex]
 & & \dsp{ +\, \frac{1}{2} \lh 3 \lb_3 - 2 \lb_1 \lb_2 \rh \bPs_+ \Ps_+\, 
 \bPs_- \Ps_- + \frac{1}{2}\, g_{z\bz} \bPs_+ \gam^{\mu} \Ps_-\, 
 \bOg_+ \gam_{\mu} \Og_- ,}
\nonumber \\[2ex]
T_{\mu\nu}(C= 0) &=& V_{\mu} V_{\nu} + \frac{1}{4}\,\bPs_+ \lh \gam_{\mu} 
 \stackrel{\leftrightarrow}{\der}_{\nu} + \gam_{\nu} 
 \stackrel{\leftrightarrow}{\der}_{\mu} \rh \Ps_- - \frac{i}{4}\,\lb_1\,
\bPs_+ \lh \gam_{\mu} V_{\nu} + \gam_{\nu} V_{\mu} \rh \Ps_-
\nonumber \\[2ex] 
& &- g_{\mu\nu} \left[ \frac{1}{2}\,V^2 + \frac{1}{2}\,\bPs_+ \stackrel{
\leftrightarrow}{\sder} 
 \Ps_- + \,\frac{\lb_2}{4}\bPs_+ \Ps_+ \bPs_- \Ps_- \right].
\label{5.7} 
\end{eqnarray}
The physical interpretation of these equations is implicit in their bosonic 
terms. For a hydrodynamical current 
\be
V_{\mu} = \rg u_{\mu} \hs{1} \Rightarrow \hs{1} V_{\mu}^2 = - \rg^2.
\label{5.8}
\ee
The bosonic part of the first equation (\ref{5.7}) becomes 
\be 
\rg^2 = - \frac{4}{\lb_1}\, g_{z\bz} \der \bz \cdot \der z \geq 0.
\label{5.9}
\ee
In particular, for $\lb_1 < 0$ it becomes
\be 
\rg^2 = \frac{4}{|\lb_1|}\, g_{\bz z} \lh |\mbox{\boldmath{$\nabla$}} z|^2 - 
| \dot{z} |^2 \rh,
\label{5.10}
\ee
which implies that apart from fermionic contributions, the spatial gradient 
of the complex 
scalar field determines the fluid density $\rg$. Similarly, for $\lb_1 > 0$ 
the time rate of change of $z$  determines $\rg$. With the identification
(\ref{5.8}), the bosonic part of the energy-momentum 
(\ref{5.7}) is of the form (\ref{1.1}). The 
corresponding energy and pressure densities are given by
\be
\varepsilon = p = \frac{1}{2}\rg^2 
= \frac{2}{|\lb_1|}\, g_{\bz z} \lh |\mbox{\boldmath{$\nabla$}} z|^2 - 
| \dot{z} |^2 \rh.
\label{5.11}
\ee

It is not difficult to check, that the equation for the complex scalar fields $(\bz, z)$ for $C = 0$ 
reproduce the conditions
\be 
V \cdot \der z = V \cdot \der \bz = 0, 
\label{5.11.2}
\ee
cf.\ eqs.\ (\ref{1.9}). This establishes the existence of a regime $C = 0$ in which the 
supersymmetric model allows a fluid-mechanical interpretation.

In the following, we show that the full energy-momentum tensor
(\ref{5.7}) takes the standard form (\ref{1.1}), under particular
condtions specified by 
\be
\der_\mu\Ps_\pm = \pm\,\frac{i}{2}\lb_1V_\mu\Ps_\pm 
- \frac{\lb_2}{8}\gam_\mu
\Ps_\mp\,\bPs_\pm\,\Ps_\pm,
\quad
\lb_1^2 = \lb_2
\label{anzats}.
\ee
Indeed upon substitution of these expressions into (\ref{5.7}), the
energy-momentum becomes  
\begin{eqnarray}
T_{\mu\nu} &=& W_{\mu} W_{\nu} - \frac{1}{2} g_{\mu\nu} W^2,\hs{1} W_{\mu} = 
V_\mu - \frac{i}{2}\lb_1\bPs_+\gam_\mu\Ps_-,\hs{1}\der\cdot W = 0.
\label{energy} 
\end{eqnarray}
The vanishing divergence of the current $W_\mu$ follows upon using the field 
equations
\be
\sder\Ps_\pm = \pm\,\frac{i}{2}\lb_1V\Slashed\Ps_\pm - \frac{\lb_2}{2}
\Ps_\mp\,\bPs_\pm\,\Ps_\pm,\hs{1}\der\cdot V = 0.
\label{fequations}
\ee
Therefore, we can reinterpret the $W_\mu$ as the hydrodynamical current 
$W_\mu = \rg u_{\mu}$ with the equation of state:
\begin{eqnarray}
\varepsilon &=& p = \frac{1}{2}\rg^2,\hs{1}\rg^2 = - \frac{1}{\lb_1}\Bigl[
4 g_{z\bz} \der \bz \cdot \der z + 
\frac{1}{2} \lh 3 \lb_3 - 2 \lb_1^3\rh \bPs_+ \Ps_+\, 
 \bPs_- \Ps_-\Bigl]\nonumber.
\label{identi}
\end{eqnarray}
Next, we investigate the properties of non-trivial solutions to the
Ansatz (\ref{anzats}). Since (\ref{anzats}) can be written as
\be
\cD_\mu \Ps_\pm =  
\Bigl(\der_\mu\mp\frac{i\lb_1}{2}W_\mu\Bigl)\Ps_\pm = 0, 
\ee
with the axial covariant derivative 
$\cD_\mu = \der_\mu - \frac{i\lb_1}{2}W_\mu\, \gamma_5$
using a Fierz identity, the ansatz has the formal solution 
\be
\Ps_\pm (x) = {\cal P}\exp\Bigl(\pm\frac{i\lb}{2}\int_{0}^{x}W\cdot dx\Bigl)
\Ps_\pm(0)
\label{PsSol}
\ee
where, ${\cal P}$ is the path-ordering operator. In fact, this
solution shows that the fermion bilinear 
\be 
\bar \Ps_+(x) \gam_{\mu} \Ps_-(x) = \bar \Ps_+(0) \gam_{\mu} \Ps_-(0) 
\ee
is constant. On the other hand, from (\ref{PsSol}) one infers that the
field strength associated with the covariant derivative $\cD_\mu$
vanishes. This implies that $W_\mu$ has to be pure gauge for
non--trivial fermion solutions to exist. And because the fermion
bilinear is constant, we conclude that $V_\mu$ is pure gauge. This
shows that the system in this regime is described by potential flow. 
\nl

\nit
Finally, we return to the identically conserved currents constructed from Killing potentials. 
For the simplest K\"{a}hler geometry, the complex plane, the potential is
\be
K(\bz,z) = \bz z.
\label{5.12}
\ee
The plane admits the following holomorphic Killing vectors, related to translations and rotations
in the plane:
\be
\del z = R^z(z) = \eps + i \ag z, \hs{2} \del \bz = \bR^{\bz}(\bz) = \bar{\eps} - i \ag \bz.
\label{5.13}
\ee
with the property 
\be
\del K = \eps \bz + \bar{\eps} z.
\label{5.14}
\ee
The corresponding Killing potentials are 
\be
G(\bz,z) = i \lh \bar{\eps} z - \eps \bz \rh + \ag\bz z.
\label{5.15}
\ee 
If we insert these expressions into the brackets (\ref{2.17}), (\ref{2.19}) we obtain
the transformations generated by the conserved Noether charges
\be
Q[G] = \int d^3 x\, J_0[G],
\label{5.15.1}
\ee
then we indeed reobtain the isometries:
\be \ba{l}
\del_G z = R^z(z), \hs{2} \del_G \bz = \bR^{\bz}(\bz),\\
\\
\del_G \thg = 
i ( \eps \bz - \bar{\eps} z).
  \ea 
\label{5.16}
\ee
By including the fermionic contributions to the conserved Noether charges 
(\ref{4.19}), one finds in addition to (\ref{5.16}) the transformations of 
other fields
\be
\ba{l} 
\dsp{ \del_G \eta_- = \left\{ Q[G], \eta_- \right\}^* 
 = R^z_{,z}(z)\eta_-,}\\
 \\
\dsp{ \del_G \eta_+ = \left\{ Q[G],\eta_+  \right\}^* = \bR^{\bz}_{,\bz}(\bz)
\eta_+,\hs{1}\del_G V_\mu = \left\{ Q[G], V_\mu \right\}^* = 0 }\\
 \\
\dsp{ \del_G \ps_\pm = \left\{ Q[G], \ps_\pm \right\}^* = 0,\hs{1}\del_G C = \left\{ Q[G], C 
\right\}^* = 0. }
\ea
\ee 
For the details see ref.\ \ct{tsn}.

\np
\section{Conclusions}

In this paper we have investigated the geometrical structure of
relativistic fluid mechanics and its supersymmetric
extension. 

Conventionally, a relativistic fluid with non-vanishing
vorticity is described by the Clebsch decomposition of the current. We
propose a different parameterization in terms of one real and one
complex degree of freedom. The complex degree of freedom $z$ takes 
its values on a K\"ahler manifold. In this formulation the infinite
set of conserved currents $J_\mu[G]$ of fluid dynamics are associated
with the set of functions $G$ of this complex variable $z$ and its
conjugate $\bz$. A canonical analysis using the Poisson-Dirac bracket
showed, that the closure of the algebra of conserved currents leads to
a Poisson bracket structure for these functions on the K\"ahler
manifold. In eqs.\ (\ref{2.22}) this Poisson structure is given in
terms of the inverse K\"ahler metric. 

However, the main advantage of our K\"ahler parameterization of the
fluid current is, that it allows for a rather straightforward
supersymmetric completion. The lagrangean for the supersymmetric
system (\ref{4.1}) has been expressed in terms of superfields, which
allows both an on- and off-shell supersymmetric formulation. The only
restriction which our formulation seems to impose is, that the
energy density is proportional to the square of the current density. 
Contrary to a fluid mechanical system, the supersymmetric model does
not possess an infinite set of conserved: In general, the currents
$J_\mu[G]$ are only conserved, if the function $G$ is a Killing
potential associated to an isometry of the K\"ahler manifold. However,
the theory does contain a regime, in which the infinite number of
currents are conserved. This regime arises in a supersymmetric
background in which expectation values of the fermions and some
bosonic fields are zero.  
\vs{2}

\nit
{\bf Acknowledgements} For two of us (T.S.N.\ and J.W.v.H.) this work is 
performed as part of the program FP52 of the Foundation for Fundamental 
Research of Matter (FOM). S.G.N.\ acknowledges the support of CITA and
NSERC. 

\np
\begin{appendix} 

\section{Killing vectors and their currents}

A K\"{a}hler manifold is a complex manifold with a curl-free metric; the latter 
condition implies that locally one can write the metric as the mixed second derivative
of a real scalar function $K(\bz,z)$, the K\"{a}hler potential: $g_{z\bz} = K_{,z\bz}$. 
This metric is invariant under infinitesimal holomorphic co-ordinate transformations 
$\del z = R^z(z)$, $\del \bz = \bR^{\bz}(\bz)$ if and only if
\be
R_{\bz;z} + \bR_{z;\bz} = 0. 
\label{a.1}
\ee
The solutions of these equations are holomorphic Killing vectors. The invariance of
the metric implies that the K\"{a}hler potential can only change by the real part of a 
holomorphic function: 
\be 
\del K(\bz,z) = R^z K_{,z} + \bR^{\bz} K_{,\bz} = F(z) + \bF(\bz)
\label{a.2}
\ee
From this we infer the existence of a real scalar function $G(\bz,z)$ such that
\be 
R^z K_{,z} - F = - \bR^{\bz} K_{,\bz} + \bF = i G.
\label{a.3}
\ee
As both $R^z$ and $F$ are holomorphic functions, and the metric is
covariantly constant, it follows that 
\be 
G_{,\bz} = - i g_{z\bz} R^ z = -i R_{\bz} \hs{1} \Leftrightarrow \hs{1} 
 G_{;\bz;\bz} = -i R_{\bz;\bz} = 0. 
\label{a.4}
\ee 
Because of this relation $G$ is called a Killing potential. 
The above  argument runs both ways, because first the condition $G_{;\bz;\bz} = 0$ 
can be integrated to give 
\be 
i G_{,\bz}\, g^{\bz z} = R^z(z),
\label{a.5}
\ee
a holomorphic function; and second because this implies that the function $F$ defined by
\be 
F = R^z K_{,z} - i G,
\label{a.6}
\ee
then is also holomorphic: $F_{,\bz} = 0$. Therefore under a transformation $\del z = R^z(z)$ 
and its complex conjugate the K\"{a}hler potential changes only by the real part of a 
holomorphic function, and the metric is invariant.

\end{appendix}

\np

\end{document}